\documentclass[fp,twocolumn]{jpsj3}
\usepackage{txfonts}

\title{$^{63,65}$Cu Nuclear Resonance Study of the Coupled Spin Dimers and Chains Compound Cu$_2$Fe$_2$Ge$_4$O$_{13}$}

\author{\name{Jun \surname{Kikuchi}}$^{1,}$\thanks{E-mail address: jkiku@isc.meiji.ac.jp}, \name{Shiro \surname{Nagura}}$^{1}$, \name{Kazumasa \surname{Murakami}}$^{1}$, \name{Takatsugu \surname{Masuda}}$^{2}$, and \name{G\"{u}nther J. \surname{Redhammer}}$^{3}$
}
\inst{$^{1}$Department of Physics, School of Science and Technology, Meiji University, \address{Kawasaki 214-8571}
\\
$^{2}$Institute for Solid State Physics, University of Tokyo,
\address{Kashiwa, Chiba 277-8581}\\

$^{3}$Department of Materials Engineering and Physics, University of Salzburg, 
\address{Salzburg A-5020, Austria}
}

\abst{Nuclear magnetic resonance (NMR) and nuclear quadrupole resonance (NQR) of Cu have been measured in a coupled spin dimers and chains compound Cu$_2$Fe$_2$Ge$_4$O$_{13}$. Cu NQR has also been measured in an isostructural material Cu$_2$Sc$_2$Ge$_4$O$_{13}$ including only spin dimers. Comparison of the temperature dependence of the $^{63}$Cu nuclear spin-lattice relaxation rate between the two compounds reveals that the Fe chains in Cu$_2$Fe$_2$Ge$_4$O$_{13}$ do not change a spin gap energy of the Cu dimers from that in Cu$_2$Sc$_2$Ge$_4$O$_{13}$, contributing additionally to the relaxation rate at the Cu site. A modestly large internal field of 3.39 T was observed at the Cu site in the antiferromagnetic state of Cu$_2$Fe$_2$Ge$_4$O$_{13}$ at 4.2 K, which is partly because of quantum reduction of the ordered moment of a Cu atom. The internal field and the ordered moment of Cu are noncollinear due to large anisotropy of the hyperfine interaction at the Cu site. A model analysis of the internal field based on the fourfold planar coordination of Cu suggests that a 3$d$ hole of the Cu$^{2+}$ ion is mainly in the $d(x^2-y^2)$ orbital state. 
}

\kword{NMR, NQR, Cu$_2$Fe$_2$Ge$_4$O$_{13}$, Cu$_2$Sc$_2$Ge$_4$O$_{13}$, spin dimer, spin chain, nuclear spin-lattice relaxation}

\begin{document}
\maketitle

\section{Introduction}

Effects of quantum fluctuation on the magnetic properties of low-dimensional spin systems appear in a variety of ways depending on dimensionality of a system, a spin quantum number, geometry of interactions, and so on. Systems like spin dimers, ladders, integer spin chains have a disordered ground state with a gap in the spin excitation spectrum, being characterized by a short range spin correlation. This state is robust for external perturbations such as three-dimensional interactions and an addition of spin defects, and is not easily transformed into a state with long-range magnetic order. In contrast, systems with gapless excitations such as half-integer spin chains are more easily driven to have long-range order, reflecting a quantum critical nature of the ground state which has a divergently large spin correlation length. 

\begin{figure}[b]
\centerline{\includegraphics[width=85mm]{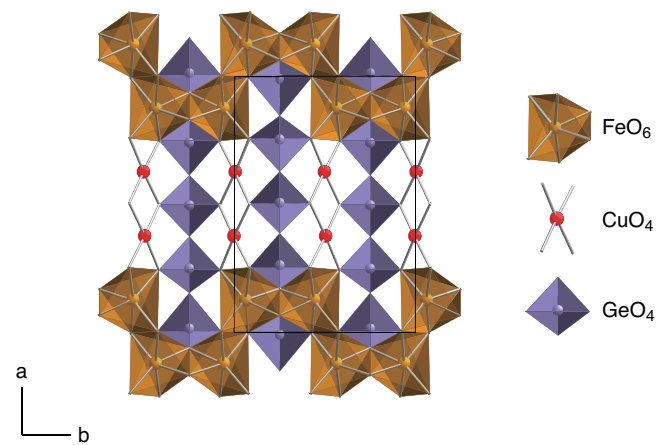}}
\caption{(Color online) The crystal structure of Cu$_2$Fe$_2$Ge$_4$O$_{13}$ projected onto the $ab$ plane. The unit cell is shown by a solid line.}
\label{fig:structure}
\end{figure}

Recently, fascinating model systems combining both types of spin networks have been realized experimentally and have attracted much attention expecting exotic phases and  transitions between them.\cite{zheludev98,masuda04,hase05,hamasaki08} Cu$_2$Fe$_2$Ge$_4$O$_{13}$ is one of the members of such a new family of materials, comprising of quantum spin dimers of Cu$^{2+}$ ions ($S$ = 1/2) and classical spin chains of Fe$^{3+}$ ions ($S$ = 5/2). 
The crystal structure of Cu$_2$Fe$_2$Ge$_4$O$_{13}$ is shown in Fig. \ref{fig:structure}. 
Cu$_2$Fe$_2$Ge$_4$O$_{13}$ crystallizes in the monoclinic space group $P2_1/m$ with the unit cell dimensions $a$ = 12.101 \AA, $b$ = 8.497 \AA, $c$ = 4.869 \AA, and $\beta$ = 96.131$^\circ$ at room temperature.\cite{masuda04,masuda03} 
The structure consists of three groups of metal-oxygen polyhedra: crankshaft-shaped chains of edge-sharing FeO$_6$ octahedra running in the $b$ direction, dimers of edge-sharing CuO$_4$ squares bridging the FeO$_6$ chains along the $a$ axis, and GeO$_4$ tetrahedra separating the chain-dimer  blocks stacked in the $c$ direction. 
It can also be viewed as being build from blocks of the famous spin-Peierls compound CuGeO$_3$\cite{hase93} and the Fe oxide chains. 

Cu$_2$Fe$_2$Ge$_4$O$_{13}$ orders antiferromagnetically below the N\'{e}el temperature $T_\mathrm{N}$ = 39 K.\cite{masuda04} The magnetic structure is roughly collinear with slight canting of ordered moments between the Cu and Fe sublattices. 
The ordered moments of Cu and Fe are estimated to be 0.38 and 3.62 $\mu_\mathrm{B}$ at 1.5 K, respectively. There are large and moderate reductions of the Cu and Fe moments, indicating strong quantum fluctuations. 
In the paramagnetic state, a temperature dependence of the bulk magnetic susceptibility is roughly reproduced by a sum of contributions of isolated $S$ = 5/2 chains and $S$ = 1/2 dimers. This has rendered support to a naive picture of Cu$_2$Fe$_2$Ge$_4$O$_{13}$ being a bicomponent magnet with weakly coupled spin chains and dimers. 

The dispersion relations of magnetic excitations have been determined by inelastic neutron scattering experiments.\cite{masuda04,masuda05,masuda07} The excitation spectrum consists of two distinct contributions: the dispersive modes at low energies below about 10 meV and the high energy mode around 24 meV with only weak dispersion. These modes are assigned respectively to spin wave excitations of the Fe chains and singlet-triplet excitations of the Cu dimers. This assignment has been supported by a recent theoretical calculation of the magnetic excitation spectrum.\cite{matsumoto10a,matsumoto10b} 
A set of exchange parameters has also been evaluated from analyses of the dispersion relations. The dominant interactions are all antiferromagnetic consistent with the magnetic structure and are obtained as follows: the intradimer Cu-Cu interaction $J_\mathrm{Cu}$ = 24 meV, the intrachain Fe-Fe interaction $J_\mathrm{Fe}$ = 1.60 meV, the interchain Fe-Fe interaction $J_\mathrm{Fe}^\prime$ = 0.12 meV in the $c$ direction, and the Cu-Fe interaction $J_\mathrm{Cu-Fe}$ = 2.54 meV. 

One of the peculiarities of the magnetic excitations in Cu$_2$Fe$_2$Ge$_4$O$_{13}$ is that the Cu dimers and the Fe chains contribute almost independently to the excitation spectrum at different energies, despite the fact that the coupling $J_\mathrm{Cu-Fe}$ between the Cu and Fe subsystems is comparable to the intrachain coupling $J_\mathrm{Fe}$. 
Eventually, the low energy modes may consistently be explained by a spin wave theory for the Fe chains with the effective Fe-Fe coupling $J_\mathrm{eff}$ = 0.09 meV via the Cu dimer as well as the interchain coupling $J_\mathrm{Fe}^\prime$ in the $c$ direction.\cite{masuda07}. 
Masuda \textit{et al}. have pointed out that such a separation of energy scales of the magnetic excitations between the two subsystems arises from a hierarchy of the exchange interactions $J_\mathrm{Cu}$ $\gg$ $J_\mathrm{Fe}$, $J_\mathrm{Cu-Fe}$.\cite{masuda04,masuda05} 
The suppression of the effective Fe-Fe interaction $J_\mathrm{eff}$ is interpreted as resulting from a nonmagnetic nature of the Cu dimers which have an excitation energy much larger than the relevant energy scales of the Fe chains. 

Although the magnetic excitation spectrum of Cu$_2$Fe$_2$Ge$_4$O$_{13}$ may be understood as being contributed by effectively-decoupled Cu dimers and Fe chains, it is highly desirable to identify the modes more definitely using a microscopic probe complementary to the neutron scattering. 
Nuclear magnetic resonance (NMR) and nuclear quadrupole resonance (NQR) are best suited for such a purpose, because they can probe local spin state and excitations quite sensitively. In this paper, we report on the results of 
NMR and NQR measurements of $^{63}$Cu and $^{65}$Cu nuclei (both having nuclear spin $I$ = 3/2) in Cu$_2$Fe$_2$Ge$_4$O$_{13}$ and in an isostructural compound Cu$_2$Sc$_2$Ge$_4$O$_{13}$ in which Fe chains are all replaced by nonmagnetic Sc chains.\cite{redhammer04} 
Cu$_2$Sc$_2$Ge$_4$O$_{13}$ contains Cu dimers which are magnetically well isolated from each other and have a spin gap of about 24 meV.\cite{masuda06,lue07,jkiku10} The room-temperature lattice parameters are $a$ = 12.336 \AA, $b$ = 8.703 \AA, $c$ = 4.888 \AA, and $\beta$ = 95.74$^\circ$. \cite{redhammer04} 
Our measurements of the $^{63}$Cu nuclear spin-lattice relaxation rate in the two compounds reveal that Cu$^{2+}$ and Fe$^{3+}$ spin fluctuations in Cu$_2$Fe$_2$Ge$_4$O$_{13}$ are decoupled to contribute additionally to the relaxation rate at the Cu site. The dynamics of Cu$^{2+}$ spins in Cu$_2$Fe$_2$Ge$_4$O$_{13}$ is characterized by a spin gap identical to that in Cu$_2$Sc$_2$Ge$_4$O$_{13}$, which allows us to unambiguously identify the high energy mode as magnetic excitations of the Cu dimers. We also find from the relation between directions of the ordered moment of Cu and the internal field at the Cu nuclear site that the electronic orbital of a Cu$^{2+}$ ion in Cu$_2$Fe$_2$Ge$_4$O$_{13}$ has predominantly a $d(x^2-y^2)$ character. 

In \S\ref{sec:expt} we describe the experimental procedures. The results are presented and analyzed in \S\ref{sec:result}, where the characteristics of Cu$^{2+}$ and Fe$^{3+}$ spin fluctuations in Cu$_2$Fe$_2$Ge$_4$O$_{13}$ are deduced from the nuclear spin-lattice relaxation rate of Cu. 
The internal magnetic field and the electric field gradients at the Cu site are also determined in \S\ref{sec:result} through an analysis of the Cu NMR spectrum in the antiferromagnetic state. In \S\ref{sec:disc} we discuss an electronic state of the Cu$^{2+}$ ion in the Fe compound and the nuclear spin-lattice relaxation rate at high temperatures in both the Fe and Sc compounds. Summary of the paper will be given in the final section \ref{sec:summary}. 

\section{Experiments}
\label{sec:expt}

Single crystals of Cu$_2$Fe$_2$Ge$_4$O$_{13}$ were grown by the floating zone method\cite{masuda04}. The dimension of the crystal used in the experiments was $5\times 3.5\times 3.2$ mm$^3$. A polycrystalline sample prepared by the solid-state reaction technique\cite{redhammer04} was used for measurements in Cu$_2$Sc$_2$Ge$_4$O$_{13}$. NQR and NMR experiments were performed with a phase-coherent type pulsed spectrometer 
at zero external magnetic field. Standard $\pi/2$-$\pi$ two-pulse sequence was used to excite spin-echo signals. The Cu NQR spectra were taken by Fourier-transforming the spin-echo signal with the step-sum technique.\cite{clark95} The nuclear spin-lattice relaxation rate was measured by the inversion recovery method. The Cu NMR spectrum in the antiferromagnetic state of Cu$_2$Fe$_2$Ge$_4$O$_{13}$ was taken by recording an integrated intensity of the spin-echo signal point by point in the frequency range from 3.6 to 100 MHz.

\begin{figure}[t]
\centerline{\includegraphics[width=95mm]{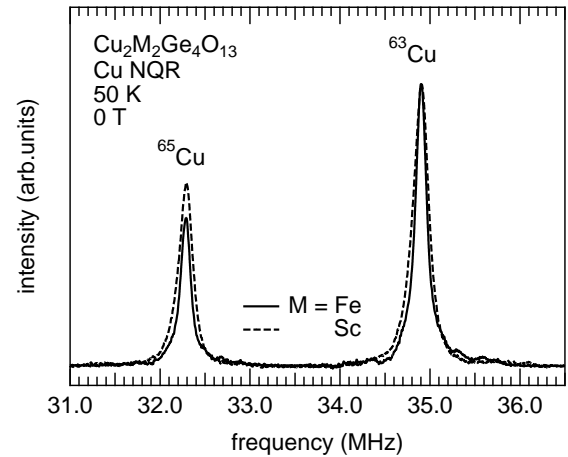}}
\caption{Cu NQR spectra in Cu$_2$Fe$_2$Ge$_4$O$_{13}$ (solid line) and Cu$_2$Sc$_2$Ge$_4$O$_{13}$ (dashed line) at 50 K. Intensities are normalized at the peak values.}
\label{fig:NQR_spectra}
\end{figure}

\section{Results and Analyses}
\label{sec:result}
\subsection{Cu NQR spectrum}

A pair of sharp resonance lines was observed at zero field in the paramagnetic state of Cu$_2$Fe$_2$Ge$_4$O$_{13}$ (Fig. \ref{fig:NQR_spectra}).  We identified the signal as Cu NQR from a unique crystallographic site because a ratio between the peak frequencies agrees with that between the electric quadrupole moments of $^{63}$Cu and $^{65}$Cu, $^{63}Q/^{65}Q=1.081$. This is consistent with the crystal structure in which all the Cu atoms occupy equivalent positions.\cite{masuda03} The $^{63}$Cu NQR frequency determined from the peak position of the resonance line is 34.90 MHz at 50 K.
The lines are broadened by inhomogeneous distribution of the electric field gradient (EFG) at the Cu site. The full width at half maximum (FWHM) is 149 kHz at 50 K for $^{63}$Cu. 

In Fig. \ref{fig:NQR_spectra} the Cu NQR spectrum in Cu$_2$Sc$_2$Ge$_4$O$_{13}$ is also shown for comparison. The $^{63}$Cu NQR frequency is 34.91 MHz at 50 K, coinciding with that in Cu$_2$Fe$_2$Ge$_4$O$_{13}$ at the same temperature within experimental accuracies. On the other hand, the line is broader in Cu$_2$Sc$_2$Ge$_4$O$_{13}$ with the FWHM of 206 kHz at 50 K for $^{63}$Cu. This indicates that the local structure around Cu is more homogeneous in the Cu$_2$Fe$_2$Ge$_4$O$_{13}$ crystal specimen. 

\begin{figure}[t]
\centerline{\includegraphics[width=80mm]{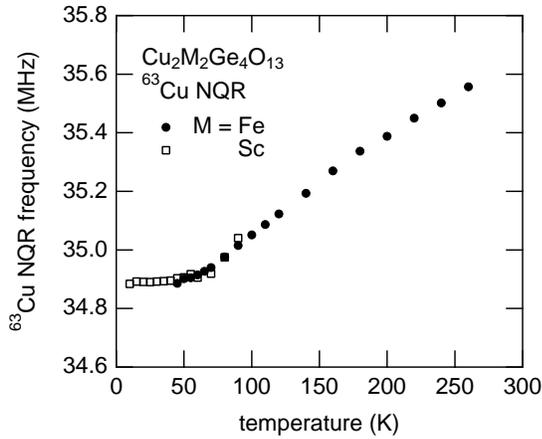}}
\caption{Temperature dependence of the $^{63}$Cu NQR frequency in Cu$_2$Fe$_2$Ge$_4$O$_{13}$ (solid circles) and Cu$_2$Sc$_2$Ge$_4$O$_{13}$ (open squares).}
\label{fig:NQR_Tdep}
\end{figure}

The temperature dependence of the $^{63}$Cu NQR frequency $^{63}\nu_\mathrm{NQR}$ in the two compounds is almost identical as shown in Fig. \ref{fig:NQR_Tdep}, although the data for Cu$_2$Sc$_2$Ge$_4$O$_{13}$ are limited up to 90 K due to fast nuclear spin relaxation processes. $^{63}\nu_\mathrm{NQR}$ in Cu$_2$Sc$_2$Ge$_4$O$_{13}$ is independent of temperature below about 50 K. As the temperature is raised, $^{63}\nu_\mathrm{NQR}$ continues to increase up to 260 K in Cu$_2$Fe$_2$Ge$_4$O$_{13}$. This is opposite to a tendency of the EFG to decrease with increasing temperature due to lattice thermal expansion. 
Such a behavior of the Cu NQR frequencies has been reported in some Cu oxides including high-$T_\mathrm{C}$ superconductors, and is attributed to a dominant contribution of on-site electronic orbitals to the EFG which is different in sign from the contribution of surrounding ions.\cite{shimizu93}

\subsection{Nuclear spin-lattice relaxation}
\label{sec:T1}

The nuclear spin-lattice relaxation rate $1/T_{1,\mathrm{NQR}}$ at the Cu site in Cu$_2$Fe$_2$Ge$_4$O$_{13}$ was determined as a time constant of exponential recovery of the nuclear magnetization. Non-exponential recovery was observed in Cu$_2$Sc$_2$Ge$_4$O$_{13}$ below 45 K due to contributions of paramagnetic impurities. $1/T_{1,\mathrm{NQR}}$ was then determined by fitting the nuclear magnetization $M(t)$ to the stretched exponential form which incorporates both the impurity and intrinsic relaxations,
\begin{align}
	M(t)=M_\infty[1-p_0\exp{(-\sqrt{t/\tau_c}-t/T_{1,\mathrm{NQR}})}],
\end{align}
where $M_\infty$ is the magnetization in thermal equilibrium, $p_0$ is a parameter describing a degree of inversion, and $1/\tau_c$ is the relaxation rate induced by paramagnetic impurities\cite{mchenry72}. The impurity relaxation rate $1/\tau_c$ is almost temperature independent and takes a value of about 1 $\mathrm{s^{-1}}$ for $^{63}$Cu. In both compounds the isotopic ratio of $1/T_{1,\mathrm{NQR}}$ between $^{63}$Cu and $^{65}$Cu agrees with the ratio of the square of the gyromagnetic ratios $(^{63}\gamma/^{65}\gamma)^2$ within experimental accuracies. The relaxation process is thus magnetic in origin. 

\begin{figure}[t]
\centerline{\includegraphics[width=94mm]{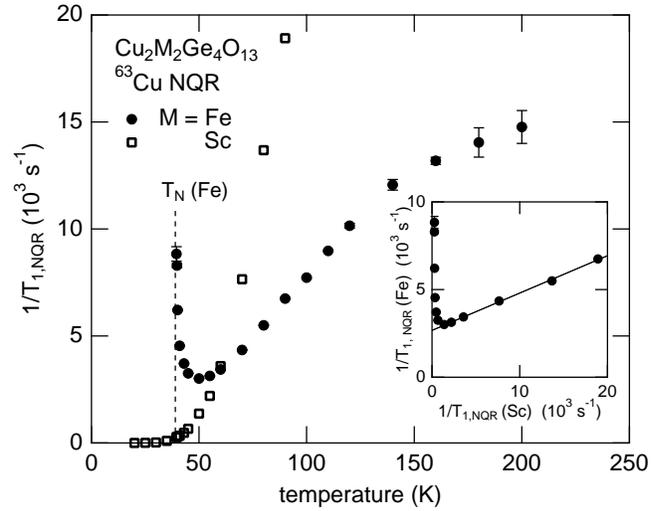}}
\caption{
Temperature dependences of the nuclear spin-lattice relaxation rate $1/T_{1,\mathrm{NQR}}$ at the $^{63}$Cu site in Cu$_2$Fe$_2$Ge$_4$O$_{13}$ (solid circles) and Cu$_2$Sc$_2$Ge$_4$O$_{13}$ (open squares). Dashed line stands for $T_\mathrm{N}$ of Cu$_2$Fe$_2$Ge$_4$O$_{13}$. The inset is a plot of $1/T_{1,\mathrm{NQR}}\mathrm{(Fe)}$ versus $1/T_{1,\mathrm{NQR}}\mathrm{(Sc)}$ with temperature the implicit parameter. Solid line is a fit of $1/T_{1,\mathrm{NQR}}\mathrm{(Fe)}$ above 50 K.
}
\label{fig:T1}
\end{figure}

Figure \ref{fig:T1} shows the temperature dependence of $1/T_{1,\mathrm{NQR}}$ at the Cu site in the two compounds. The relaxation rate $1/T_{1,\mathrm{NQR}}(\mathrm{Fe})$ in Cu$_2$Fe$_2$Ge$_4$O$_{13}$ increases gradually with increasing temperature well above $T_\mathrm{N}$, exhibiting a tendency to saturate at high temperatures. 
On the other hand, it increases divergently on approaching $T_\mathrm{N}$. This signals critical slowing down of electronic spin fluctuations toward long-range magnetic order. 

\begin{figure}[t]
\centerline{\includegraphics[width=82mm]{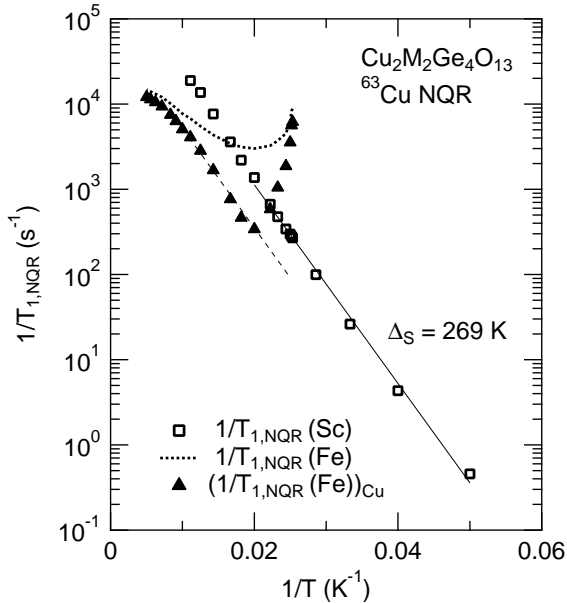}}
\caption{
Nuclear spin-lattice relaxation rate $1/T_{1,\mathrm{NQR}}\mathrm{(M)}$ at the $^{63}$Cu site in Cu$_2$Sc$_2$Ge$_4$O$_{13}$ (M = Sc; solid squares) and Cu$_2$Fe$_2$Ge$_4$O$_{13}$ (M = Fe; dotted line) plotted against $1/T$. Solid line is a fit of $1/T_{1,\mathrm{NQR}}\mathrm{(Sc)}$ to the activation law with a spin gap $\Delta_\mathrm{S}$ = 269 K. Solid triangles are the contribution of Cu$^{2+}$ spins $(1/T_{1,\mathrm{NQR}}\mathrm{(Fe)})_\mathrm{Cu}$ to $1/T_{1,\mathrm{NQR}}\mathrm{(Fe)}$. Dashed line is drawn as a guide to the eyes to demonstrate an activated behavior of $(1/T_{1,\mathrm{NQR}}\mathrm{(Fe)})_\mathrm{Cu}$ with the same spin gap as that of $1/T_{1,\mathrm{NQR}}\mathrm{(Sc)}$.
}
\label{fig:T1_actv}
\end{figure}

As a clear contrast, a monotonous and rapid decrease of the relaxation rate $1/T_{1,\mathrm{NQR}}(\mathrm{Sc})$ in Cu$_2$Sc$_2$Ge$_4$O$_{13}$ was observed at low temperatures. This reflects a nonmagnetic ground state with a finite spin gap as expected for spin dimers. Figure \ref{fig:T1_actv} shows $1/T_{1,\mathrm{NQR}}(\mathrm{Sc})$ as a function of inverse temperature $1/T$ which evidences a thermally-activated behavior of $1/T_{1,\mathrm{NQR}}(\mathrm{Sc})$.
By fitting the data below 50 K to the formula $1/T_{1,\mathrm{NQR}}(\mathrm{Sc}) = c\exp(-\Delta_\mathrm{S}/T)$, we determined the constant $c$ and the spin gap $\Delta_\mathrm{S}$ to be $(2.3\pm 0.1)$ $\times$ 10$^5$ s$^{-1}$ and $\Delta_\mathrm{S}$ = $269\pm 10$ K, respectively. The value of $\Delta_\mathrm{S}$ agrees well with those estimated from susceptibility, $^{45}$Sc-NMR, and neutron scattering measurements.\cite{masuda06,lue07}

It is instructive to compare the temperature dependences of $1/T_{1,\mathrm{NQR}}(\mathrm{Fe})$ and $1/T_{1,\mathrm{NQR}}(\mathrm{Sc})$ to get insight into how the low-energy magnetic excitations of Cu dimers are affected by Fe chains in Cu$_2$Fe$_2$Ge$_4$O$_{13}$. 
The inset of Fig. \ref{fig:T1} is a plot of $1/T_{1,\mathrm{NQR}}(\mathrm{Fe})$ versus $1/T_{1,\mathrm{NQR}}(\mathrm{Sc})$ with temperature the implicit parameter. 
It is clear that $1/T_{1,\mathrm{NQR}}(\mathrm{Fe})$ scales with $1/T_{1,\mathrm{NQR}}(\mathrm{Sc})$ except at temperatures very close to $T_\mathrm{N}$.
Above 50 K, $1/T_{1,\mathrm{NQR}}(\mathrm{Fe})$ is expressed as
\begin{align}\label{eq:T1scale}
	\frac{1}{T_{1,\mathrm{NQR}}(\mathrm{Fe})}=a+b\frac{1}{T_{1,\mathrm{NQR}}(\mathrm{Sc})},
\end{align}
where $a$ and $b$ are temperature-independent constants. The values of $a$ and $b$ were determined by the least-square fitting to be $(2.67\pm 0.05)$ $\times$ $10^3$ s$^{-1}$ and 0.213 $\pm$ 0.004, respectively. Deviation from the scaling near $T_\mathrm{N}$ is due to critical slowing down.

Equation (\ref{eq:T1scale}) demonstrates two important characteristics of $1/T_{1,\mathrm{NQR}}(\mathrm{Fe})$: first, it has a temperature-independent contribution which is absent in Cu$_2$Sc$_2$Ge$_4$O$_{13}$, and second, the temperature-dependent part is qualitatively the same as the temperature dependence of $1/T_{1,\mathrm{NQR}}(\mathrm{Sc})$ except near $T_\mathrm{N}$. The temperature-independent term $a$ should be interpreted as a contribution of Fe$^{3+}$ spin fluctuations to $1/T_{1,\mathrm{NQR}}(\mathrm{Fe})$, because it is the term that remains as $1/T_{1,\mathrm{NQR}}(\mathrm{Sc})$ $\rightarrow$ 0. The second term of eq. (\ref{eq:T1scale}) is regarded as the relaxation rate resulting from Cu$^{2+}$ spin fluctuations under the influence of the Fe chains. 

It is worth noting that the contribution of Fe$^{3+}$ spins is additional to that of Cu$^{2+}$ spins. 
This means that the Fe$^{3+}$ and Cu$^{2+}$ spins contribute independently to $1/T_{1,\mathrm{NQR}}(\mathrm{Fe})$. The Fe$^{3+}$ spins affect the contribution of Cu$^{2+}$ spins only through a numerical factor $b$ in eq. (\ref{eq:T1scale}), without changing 
a thermally-activated character of Cu$^{2+}$ spin fluctuations. 
In Fig. \ref{fig:T1_actv} we plotted the contribution of Cu$^{2+}$ spins to $1/T_{1,\mathrm{NQR}}(\mathrm{Fe})$ evaluated as
\begin{align}\label{eq:T1Tdep}
	\left(\frac{1}{T_{1,\mathrm{NQR}}(\mathrm{Fe})}\right)_\mathrm{Cu}&=\frac{1}{T_{1,\mathrm{NQR}}(\mathrm{Fe})}-a
\end{align}
against $1/T$. 
Raw values of $1/T_{1,\mathrm{NQR}}(\mathrm{Fe})$ were also plotted for comparison. 
Clearly, $(1/T_{1,\mathrm{NQR}}(\mathrm{Fe}))_\mathrm{Cu}$ parallels $1/T_{1,\mathrm{NQR}}(\mathrm{Sc})$ above about 50 K. This indicates that the temperature dependence of $(1/T_{1,\mathrm{NQR}}(\mathrm{Fe}))_\mathrm{Cu}$ is characterized by a spin gap identical to that of $1/T_{1,\mathrm{NQR}}(\mathrm{Sc})$. 

All these observations are consistent with the results of the neutron scattering experiments\cite{masuda04,masuda05,masuda07}. 
In Cu$_2$Sc$_2$Ge$_4$O$_{13}$, a singlet-triplet excitation mode of Cu dimers is observed at high energies around 24 meV corresponding to the intradimer exchange interaction.\cite{masuda06} 
This mode is persistent in Cu$_2$Fe$_2$Ge$_4$O$_{13}$ at the same energy and is associated with  Cu dimers. 
Our result of $(1/T_{1,\mathrm{NQR}}(\mathrm{Fe}))_\mathrm{Cu}$ confirms this assignment. 
The gapped excitations of Cu dimers is thus left unchanged by Fe chains. 
On the other hand, excitation modes at low energies below about 10 meV are viewed as spin waves propagating only within the Fe subsystem. 
The low-energy modes contribute to the excitation spectrum independently of the dimer mode   
as if the Cu dimers have nothing to do with the dynamics of the Fe chains. 
The fact that the Fe$^{3+}$ spins contribute independently of the Cu$^{2+}$ spins to $1/T_{1,\mathrm{NQR}}(\mathrm{Fe})$ 
must be related to this feature of the excitation spectrum which would result from a large difference of relevant energy scales between the two subsystems.

While the temperature dependence is governed by a spin gap which is common to both compounds, 
the absolute values of $1/T_{1,\mathrm{NQR}}(\mathrm{Sc})$ and $(1/T_{1,\mathrm{NQR}}(\mathrm{Fe}))_\mathrm{Cu}$ are scaled by the numerical factor $b$.  
This factor is determined primarily by a ratio of the hyperfine coupling constant at the Cu site between the two compounds. 
Likewise, the value of $a$ contains various information such as the dynamics of Fe$^{3+}$ spins and the hyperfine interaction between the Cu and Fe sites. 
We will return to these subjects in the next section.

\subsection{Cu NMR spectrum in the antiferromagnetic state}

We observed zero-field resonance of Cu nuclei in the antiferromagnetic state of Cu$_2$Fe$_2$Ge$_4$O$_{13}$. The lines are largely shifted from the NQR frequencies in the paramagnetic state, indicating appearance of internal magnetic field at the Cu site due to ordering of Cu and Fe magnetic moments.  

\begin{figure}[t]
\centerline{\includegraphics[width=100mm]{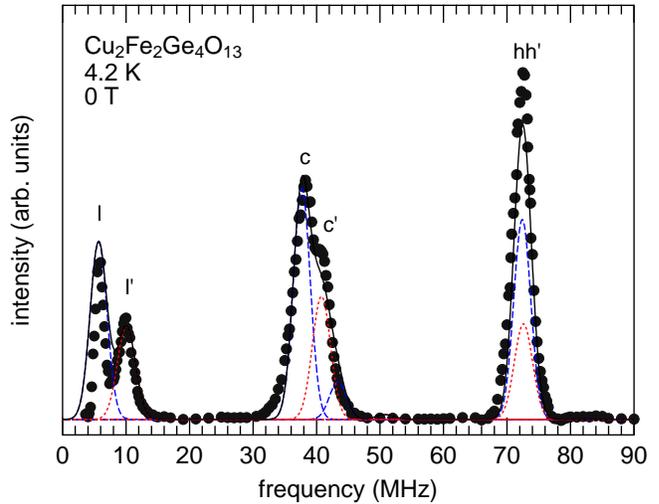}}
\caption{
(Color online) Cu NMR spectrum in the antiferromagnetic state of Cu$_2$Fe$_2$Ge$_4$O$_{13}$ at 4.2 K and 0 T. Solid line is a simulation of the spectrum based on the parameters given in the text. Dashed and dotted lines are the contributions of $^{63}$Cu and $^{65}$Cu to the spectrum, respectively. Gaussian distribution of the internal field with the FWHM of 3.1 MHz for $^{63}$Cu was assumed to simulate the spectrum.
}
\label{fig:AFNR}
\end{figure}

The spectrum is shown in Fig. \ref{fig:AFNR}. 
At 4.2 K and 0 T, three groups of resonance lines, $ll^\prime$, $cc^\prime$ and $hh^\prime$, are observed in the frequency range from 3.6 to 80 MHz. 
Such a spectrum may be interpreted as a superposition of the NMR lines of the two isotopes $^{63}$Cu and $^{65}$Cu split by the electric quadrupole interaction. 
Indeed, it is confirmed from the following analysis that the lines $c$ and $c^\prime$ correspond respectively to the central transitions of $^{63}$Cu and $^{65}$Cu, whereas the lines $l$ and $h$ ($l^\prime$ and $h^\prime$) are the quadrupole satellites of $^{63}$Cu ($^{65}$Cu). 
The overlap of the high-frequency satellites $h$ and $h^\prime$ is the reason for the reduced number (five rather than six) of the resolved peaks.
The peak frequencies and the assignment of the lines are summarized in Table \ref{tb:freq}.

\begin{table}[t]
    \caption{Observed peak frequencies and assignment of the resonance lines of the Cu NMR spectrum. Frequencies calculated using a set of parameters given in the text are also shown for comparison. Calculated intensities are normalized to the $^{63}$Cu central line $c$.}
    \label{tb:freq}
    \begin{tabular}{ccccc} \hline
	 & line & $\nu_\mathrm{obs}$ (MHz) & $\nu_\mathrm{cal}$ (MHz) & $I_\mathrm{cal}$\\ \hline
	 & $l$ & \hspace{5pt}$5.8\pm 0.2$ & \hspace{5pt}5.67 & 0.768\\
	$^{63}$Cu & $c$ & $37.8\pm 0.2$ & 37.68 & 1\\
	 & $d$ & \hspace{4pt}--- & 43.35 & 0.162\\
	 & $h$ & $72.5\pm 0.3$ & 72.40 & 0.866\\ \hline
	 & $l^\prime$ & \hspace{5pt}$9.8\pm 0.3$ & 10.05 & 0.430\\
	$^{65}$Cu & $c^\prime$ & $40.8\pm 0.2$ & 40.78 & 0.571\\
	 & $d^\prime$ & \hspace{4pt}--- & 50.82 & 0.020\\ 
	 & $h^\prime$ & $72.5\pm 0.3$ & 72.59 & 0.443\\ \hline
    \end{tabular}
\end{table}

NMR frequencies in the antiferromagnetic state under zero external field are generally determined by the Hamiltonian consisting of both the Zeeman and quadrupole interaction terms $\mathcal{H}_Z$ and $\mathcal{H}_Q$:
\begin{align}
	\mathcal{H}&=\mathcal{H}_Z+\mathcal{H}_Q,\label{eq:Hamil}\\
	\mathcal{H}_Z &=-\gamma\hbar\mathbf{I}\cdot\mathbf{B}_\mathrm{int}\nonumber\\
	&=-\gamma\hbar B_\mathrm{int}\left[I_Z\cos\theta+\frac{1}{2}\sin\theta \left(I_+ e^{-i\phi}+I_- e^{i\phi}\right)\right],\\
	\mathcal{H}_Q &=\frac{h\nu_Q}{6}\left[3I_Z^2 - \mathbf{I}^2+\frac{\eta}{2}\left(I_+^2+I_-^2\right)\right]\mathrm{~with}\nonumber\\
	\nu_Q&=\frac{3eQV_{ZZ}}{2I(2I-1)h}.
\label{eq:hamiltonian}
\end{align}
Here we refer to the principal frame of the EFG tensor.
$\gamma$ is the gyromagnetic ratio of a nucleus, 
$\mathbf{B}_\mathrm{int}$ is the internal field at the nuclear site, $Q$ is the nuclear quadrupole moment, $V_{ZZ}$ is the $Z$ component of the EFG tensor at the nucleus, and $\eta$ is the asymmetry parameter of the EFG defined as $\eta$ = $(V_{XX}-V_{YY})/V_{ZZ}$ with $V_{\mu\mu}$ = $\partial^2V/\partial \mu^2$ ($\mu=X,Y,Z$). 
We take $|V_{XX}|$ $\leq$ $|V_{YY}|$ $\leq$ $|V_{ZZ}|$ by convention so that 0 $\leq$ $\eta$ $\leq$ 1.
Polar and azimuth angles $\theta$ and $\phi$ specify the orientation of $\mathbf{B}_\mathrm{int}$ with respect to the principal axes of the EFG tensor.

In order to determine the five parameters $B_\mathrm{int}$, $\theta$, $\phi$, $\nu_Q$ and $\eta$ which best reproduce the observed resonance frequencies, we numerically diagonalized the Hamiltonian eq. (\ref{eq:Hamil}) and calculated the resonance frequencies by adjusting parameter values. 
This is because in the present case the Zeeman and quadrupole interactions are comparable in magnitude, so that the perturbation theory cannot give resonance frequencies with desired accuracy.\cite{fn:perturbation}

In searching for the parameter values, we used the value $^{63}\nu_\mathrm{NQR}$ = 34.90 MHz at 50 K as a value at 4.2 K, because the NQR frequency $\nu_\mathrm{NQR}$ in Cu$_2$Fe$_2$Ge$_4$O$_{13}$ is considered to be temperature independent below 50 K judging from the temperature dependence of $^{63}\nu_\mathrm{NQR}$ in Cu$_2$Sc$_2$Ge$_4$O$_{13}$ shown in Fig. \ref{fig:NQR_Tdep}. Since $\nu_\mathrm{NQR}$ for nuclei with $I=3/2$ is given as 
\begin{align}\label{eq:fNQR}
	\nu_\mathrm{NQR}=\nu_Q\sqrt{1+\frac{\eta^2}{3}},
\end{align}
$^{63}\nu_Q$ can be determined from $^{63}\nu_\mathrm{NQR}$ by setting $\eta$. $^{65}\nu_Q$ is then set by $^{63}\nu_Q$ using the isotopic ratio of the quadrupole moments as $^{65}\nu_Q$ = $^{63}\nu_Q\,(^{65}Q/^{63}Q)$. Values to be determined then reduce to four; $B_\mathrm{int}$, $\theta$, $\phi$ and $\eta$ which are common to both isotopes.

\begin{figure}[t]
\centerline{\includegraphics[width=63mm]{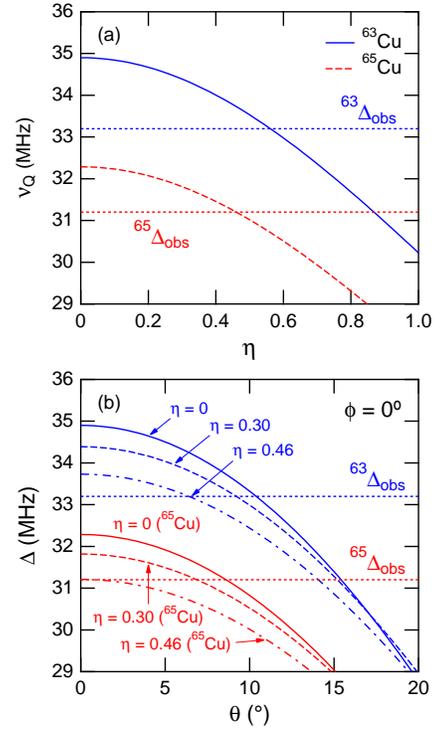}}
\caption{
(Color online) (a) Comparison of the observed quadrupole splittings $\Delta_\mathrm{obs}$ (dotted lines) with $\nu_Q$ calculated as a function of $\eta$ from $^{63}\nu_\mathrm{NQR}$ = 34.90 MHz. Solid and dashed lines represent $^{63}\nu_Q$ and $^{65}\nu_Q$, respectively. 
The condition $\Delta_\mathrm{obs}\leq\nu_Q$ is required for both isotopes. 
(b) Comparison of the observed quadrupole splitting $\Delta_\mathrm{obs}$ (dotted lines) with the splitting $\Delta$ calculated as a function of $\theta$ for $\eta$ = 0 (solid lines), 0.30 (dashed lines) and 0.46 (dotted dashed lines) using eqs. (\ref{eq:fNQR}) and (\ref{eq:split}) with $^{63}\nu_\mathrm{NQR}$ = 34.90 MHz and $\phi = 0^\circ$. For a given value of $\eta$, the solution of $\Delta(\theta)=\Delta_\mathrm{obs}$ yields a polar angle $\theta$ of the internal field.
}
\label{fig:param}
\end{figure}

Although the perturbation theory cannot be adopted to obtain accurate values of $B_\mathrm{int}$, $\theta$, $\phi$ and $\eta$, the results might be used to find appropriate ranges of parameter values. According to the second order perturbation theory, the quadrupole splitting $\Delta$ between the high- and low-frequency satellites of $I=3/2$ nuclei is given by\cite{abragam}
\begin{align}\label{eq:split}
	\Delta=\frac{\nu_\mathrm{h}-\nu_\mathrm{l}}{2}=\frac{\nu_Q}{2}\,(3\cos^2\theta-1+\eta\sin^2\theta\cos 2\phi),
\end{align}
where $\nu_\mathrm{h}$ and $\nu_\mathrm{l}$ are resonance frequencies of the high- and low-frequency satellites, respectively. 
$\Delta$ satisfies a relation
\begin{align}
	-\frac{1+\eta}{2}\nu_Q\leq\Delta\leq\nu_Q.
\end{align}
The upper and lower bounds of $\Delta$ correspond respectively to $\theta$ = $0^\circ$ and $\theta$ = $\phi$ = $90^\circ$. 
Therefore, $|\Delta|$ cannot exceed $\nu_Q=\nu_\mathrm{NQR}(1+\eta^2/3)^{-1/2}$. 
Figure \ref{fig:param}(a) compares the observed quadrupole splittings $^{63}\Delta_\mathrm{obs}$ = 33.35 MHz and $^{65}\Delta_\mathrm{obs}$ = 31.35 MHz with $^{63}\nu_Q$ and $^{65}\nu_Q$ calculated from $^{63}\nu_\mathrm{NQR}$ = 34.90 MHz as a function of $\eta$ using eq. (\ref{eq:fNQR}). The conditions ${^{63}\Delta_\mathrm{obs}}$ $\leq$ ${^{63}\nu_Q}$ and ${^{65}\Delta_\mathrm{obs}}$ $\leq$ ${^{65}\nu_Q}$ are satisfied simultaneously when $\eta$ $\leq$ 0.46 which we regard as an appropriate range of $\eta$.

Next we consider a possible range of $\theta$. Figures \ref{fig:param}(b) shows plots of $\Delta$ versus $\theta$ for several values of $\eta$ calculated using eqs. (\ref{eq:fNQR}) and (\ref{eq:split}) with $^{63}\nu_\mathrm{NQR}$ = 34.90 MHz and $\phi = 0^\circ$. 
In this plot, the solution of $\Delta(\theta)$ = $\Delta_\mathrm{obs}$ yields a polar angle $\theta$ of the internal field for a given value of $\eta$. For example, the calculated splitting $\Delta$ for $^{63}$Cu crosses with $^{63}\Delta_\mathrm{obs}$ at $\theta$ = 9$^\circ$ when $\eta$ = 0.30, yielding a polar angle of 9$^\circ$ for this value of $\eta$. 
The observed splitting $^{63}\Delta_\mathrm{obs}$ ($^{65}\Delta_\mathrm{obs}$) hence corresponds to a value of $\theta$ in a range $6^\circ$ $\leq$ $\theta$ $\leq$ 11$^\circ$ (0$^\circ$ $\leq$ $\theta$ $\leq$ 9$^\circ$) depending on $\eta$\cite{fn:Delta}. The case of $\phi\neq 0^\circ$ (not shown) gives qualitatively the same results. The range 0$^\circ$ $\leq$ $\theta$ $\leq$ 11$^\circ$ would therefore be sufficient for the parameter search. 
Note that this range of $\theta$ has an important implication on the direction of $\mathbf{B}_\mathrm{int}$: the internal field is nearly parallel to the $Z$ principal axis of the EFG tensor, being tilted by at most $\sim $10$^\circ$. 
A possibility of $\theta$ $\sim$ 90$^\circ$ can safely be excluded because for $\eta$ $\leq$ 0.46 the splitting is at most $(1+\eta)\nu_Q/2$ = $0.73\nu_Q$ = 24.62 MHz for $^{63}$Cu which is much smaller than $^{63}\Delta_\mathrm{obs}$. 

The magnitude $B_\mathrm{int}$ of the internal field may roughly be evaluated from the frequency of the central transition, neglecting the second order effect of the quadrupole interaction. The resonance frequency $\nu_\mathrm{c}$ is then given as $\nu_\mathrm{c}$ = $\gamma B_\mathrm{int}$. Using the values $\nu_\mathrm{c}$ = 37.8 MHz (40.8 MHz) and $\gamma$ = 11.285 MHz/T (12.089 MHz/T) for $^{63}$Cu ($^{65}$Cu), we obtain $B_\mathrm{int}$ = 3.35 T (3.37 T) as an approximate value.

Having had appropriate ranges of the four parameters, we searched the values of $B_\mathrm{int}$, $\theta$, $\phi$ and $\eta$ which minimize a quantity $\chi^2$ = $\sum_{i}[(\nu_{i,\mathrm{obs}}-\nu_{i,\mathrm{cal}})^2/\sigma_i^2]$ by solving numerically the Hamiltonian eq. (\ref{eq:Hamil}). 
Here $\nu_{i,\mathrm{obs}}$ ($\nu_{i,\mathrm{cal}}$) is the observed (calculated) resonance frequency of the $i$-th line, $\sigma_i$ is the experimental uncertainty of $\nu_{i,\mathrm{obs}}$, and the sum runs over all the observed lines $l$, $l^\prime$, $c$, $c^\prime$, $h$ and $h^\prime$ listed in Table \ref{tb:freq}.  
Parameter ranges investigated were 3.1 $\leq$ $B_\mathrm{int}$ $\leq$ 3.6 T, $0^\circ$ $\leq$ $\theta$ $\leq$ $11^\circ$, $0^\circ$ $\leq$ $\phi$ $\leq$ $90^\circ$ and 0 $\leq$ $\eta$ $\leq$ 0.46\cite{fn:twofold}. We found that the resonance frequencies are insensitive to $\phi$ for such a small value of $\theta$. 
We therefore optimized $B_\mathrm{int}$, $\theta$ and $\eta$ by fixing $\phi$ to some values. 

We successfully reproduced the observed resonance frequencies by properly setting $B_\mathrm{int}$, $\theta$, $\phi$ and $\eta$ within the above ranges of parameters. 
The optimized values of $\theta$ and $\eta$ depend slightly on $\phi$: they change from $4.1^\circ$ and 0.283 for $\phi$ = $0^\circ$, to $3.4^\circ$ and 0.296 for $\phi$ = $90^\circ$. 
The optimized value of $B_\mathrm{int}$, on the other hand, stays constant to take a value of 3.39 T when $\phi$ is varied from $0^\circ$ to $90^\circ$. 
These sets of parameters give almost the same resonance frequencies and the minimum value $\chi^2$ = 1.7. The resulting $\nu_{i,\mathrm{cal}}$'s are listed in Table \ref{tb:freq}. 

We now wish to know what value of $\phi$ best describes the observed NMR spectrum. For that purpose we calculated the intensity of each resonance line using the eigenstates of the Hamiltonian eq. (\ref{eq:Hamil}) for the optimized sets of parameters. 
Intensity $I_{\alpha\beta}$ of the resonance line arising from transition between the states $|\alpha\rangle$ and $|\beta\rangle$ is given as $I_{\alpha\beta}$ $\propto$ $N\gamma^2\hbar^2|\langle\alpha|\mathbf{I}\cdot\mathbf{B}_1|\beta\rangle|^2$, where $N$ is the abundance of a nuclear species and $\mathbf{B}_1$ is an rf exciting field. We take a powder average of $I_{\alpha\beta}$ because the direction of $\mathbf{B}_1$ with respect to the EFG principal axes is unknown. 

\begin{figure}[t]
\centerline{\includegraphics[width=68mm]{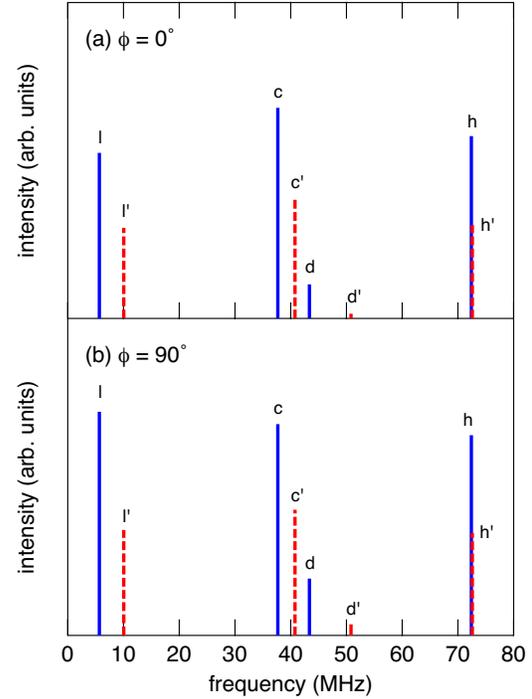}}
\caption{
(Color online) Cu NMR spectra calculated using the  sets of parameters (a) $B_\mathrm{int}$ = 3.39 T, $\eta$ = 0.283, $\theta$ = $4.1^\circ$ and $\phi$ = $0^\circ$, and (b) $B_\mathrm{int}$ = 3.39 T, $\eta$ = 0.296, $\theta$ = $3.4^\circ$ and $\phi$ = $90^\circ$. Solid (dashed) lines labeled by letters $l,c$ and $h$ ($l^\prime,c^\prime$ and $h^\prime$) represent low-frequency satellite, central, and high-frequency satellite lines of $^{63}$Cu ($^{65}$Cu), respectively. Lines $d$ and $d^\prime$ correspond respectively to transition of $^{63}$Cu and $^{65}$Cu which is forbidden in the limit of strong Zeeman interaction. Intensities are normalized to the central line $c$.
}
\label{fig:AFNR_sim}
\end{figure}

Figure \ref{fig:AFNR_sim} shows NMR spectra calculated using the optimized values of $B_\mathrm{int}$, $\theta$ and $\eta$ for $\phi=0^\circ$ and those for $\phi=90^\circ$ presented above. The most striking feature of the calculated spectra is existence of weak resonance lines $d$ and $d^\prime$ 
not observed experimentally. 
These lines correspond respectively to transition of $^{63}$Cu and $^{65}$Cu with $\Delta m=\pm 2$ which is forbidden in the limit $|\mathcal{H}_Z|\gg|\mathcal{H}_Q|$, 
and become more intense as $\phi$ varies from $0^\circ$ to $90^\circ$\cite{fn:prob}.
Of the two, the line $d$ seems strong enough to be detected by experiments. 
The most probable explanation for the lack of observation is that the line $d$ is masked by more intense lines such as $c$ and $c^\prime$. 

The line $d$ would then appear, if observed, as an asymmetrical tail in the high frequency side of the central lines $c$ and $c^\prime$ which becomes more pronounced as $\phi$ increases. 
Since we have no definite sign of such a structure in the observed spectrum, a set of parameters for $\phi$ = $0^\circ$ which yields the minimum intensity of the lines $d$ and $d^\prime$ seems best reproduce the experimental result. 
The fact that the line $l$ is more intense than the line $c$ when $\phi=90^\circ$ favors the $\phi=0^\circ$ parameter set as well, because this is opposite to the observation. 
We finally get the following set of parameters:
\begin{align}
	B_\mathrm{int} &=3.39\pm 0.01\mathrm{~T},\nonumber\\
	\theta &= 4.1^\circ\pm 0.1\mathrm{^\circ},\nonumber\\
	\phi &= 0^\circ,\\
	^{63}\nu_Q &= 34.44\pm 0.04\mathrm{~MHz},\nonumber\\
	\eta &= 0.283\pm 0.005.\nonumber
\end{align}
Errors in $B_\mathrm{int}$, $\theta$ and $\eta$ were evaluated as a range of parameters giving $\chi^2\leq 1.8$. 
The obtained value of $B_\mathrm{int}$ is modestly large for $3d$ transition metal ions and may be attributed to reduction of the ordered moment of Cu due to quantum fluctuations. 

The intensities of the resonance lines calculated using the above parameters are given in the last column of Table \ref{tb:freq}. We also simulated the NMR spectrum as a superposition of the Gauss functions based on the resonance frequencies and intensities listed in Table \ref{tb:freq}. The result is shown in Fig. \ref{fig:AFNR} for comparison with the experiments.

\section{Discussion}
\label{sec:disc}

\subsection{Internal field and $3d$ electronic orbital}
\label{sec:orbital}

In the previous section, we determined the direction as well as the magnitude of the internal field at the Cu site in Cu$_2$Fe$_2$Ge$_4$O$_{13}$. 
The internal field $\mathbf{B}_\mathrm{int}$ at the nucleus belonging to a magnetic ion comes mostly from the ordered moment $\mathbf{m}$ of that ion via the (on-site) hyperfine coupling as \begin{align}
	\mathbf{B}_\mathrm{int}=\mathsf{A}\cdot\mathbf{m}.\label{eq:Bint}
\end{align}
$\mathsf{A}$ is the hyperfine tensor which is determined by an electronic state of the magnetic ion. Therefore, the internal field carries information not only on the ordered moment but also on the electronic orbital. 
In this subsection, we will discuss the internal field and the $3d$ electronic state at the Cu site in detail, focusing on the relative orientation of the internal field and the ordered moment. 

To proceed further, we must specify the orientation of the $Z$ principal axis of the EFG tensor with respect to the crystalline axes which is as yet undetermined. 
We assume here that the $Z$ axis is normal to 
the Cu$_2$O$_2$ dimer plane based on the fourfold planar coordination of Cu, 
although there is some distortion from the ideal square planar geometry. 
This axis is also taken as the $Z$ principal axis of the hyperfine tensor. 
The angle between the $Z$ axis and the ordered moment $\mathbf{m}_\mathrm{Cu}$ of the Cu atom 
is then calculated to be 60$^\circ$ (or 120$^\circ$ for another sublattice) using the result of the neutron diffraction experiments.\cite{masuda04} 
This is a clear contrast to our observation that the internal field $\mathbf{B}_\mathrm{int}$ at the Cu site is canted scarcely from the $Z$ axis with $\theta = 4.1^\circ$. $\mathbf{B}_\mathrm{int}$ and $\mathbf{m}_\mathrm{Cu}$ are therefore noncollinear, making a large angle of about 60$^\circ$ or more between them.  

It is clear from eq. (\ref{eq:Bint}) that $\mathbf{B}_\mathrm{int}$ and $\mathbf{m}$ become noncollinear when $\mathsf{A}$ is anisotropic and $\mathbf{m}$ is parallel to none of the principal axes of $\mathsf{A}$. 
In the present case, the hyperfine interaction of the Cu$^{2+}$ ion is highly anisotropic because a $3d$ hole occupies a spatially anisotropic orbital such as $d(x^2-y^2)$ and $d(3z^2-r^2)$. Regarding the ordered moment $\mathbf{m}_\mathrm{Cu}$, it is not parallel to any of the principal axes. 
The reason why $\mathbf{m}_\mathrm{Cu}$ and $\mathbf{B}_\mathrm{int}$ at the Cu site are noncollinear is understood in this way. 

We now ask whether the anisotropic hyperfine interaction characteristic of the Cu$^{2+}$ ion is able to account for the small canting angle of $\mathbf{B}_\mathrm{int}$. 
The most stringent test would be to calculate $\mathbf{B}_\mathrm{int}$ from the experimentally-determined $\mathsf{A}$ and $\mathbf{m}_\mathrm{Cu}$. 
Unfortunately, $\mathsf{A}$ at the Cu site in Cu$_2$Fe$_2$Ge$_4$O$_{13}$ is undetermined because of difficulty in detecting NMR signals in the paramagnetic state. 
Here we present a model calculation of the dependence of $\mathbf{B}_\mathrm{int}$ on the direction of $\mathbf{m}_\mathrm{Cu}$ using $\mathsf{A}$ typical for the Cu$^{2+}$ ion.

In calculating $\mathbf{B}_\mathrm{int}$ we assume for simplicity that the Cu site has tetragonal symmetry about the $Z$ axis. 
The ground state orbital is then either $d(x^2-y^2)$ or $d(3z^2-r^2)$. 
In reality, the point symmetry at the Cu site is lower than tetragonal, 
which will cause mixing of orbitals with various symmetry and resultant deviation of the principal axes from their ideal orientations. 
Nevertheless, our simplified model well describes the observed features of $\mathbf{B}_\mathrm{int}$ if we take $d(x^2-y^2)$ as a ground state orbital. 

Now that we have axial symmetry about the $Z$ axis, 
it is sufficient to work in the plane containing $\mathbf{m}_\mathrm{Cu}$ and the $Z$ axis. We take the $X^\prime$ axis in this plane perpendicular to the $Z$ axis. Defining $\theta_\mathrm{m}$ ($\theta_\mathrm{B}$) as an angle between $\mathbf{m}_\mathrm{Cu}$ ($\mathbf{B}_\mathrm{int}$) and the $Z$ axis, we have 
\begin{align}
	\mathbf{m}_\mathrm{Cu}&=(m_\mathrm{Cu}\sin\theta_\mathrm{m},m_\mathrm{Cu}\cos\theta_\mathrm{m}),\\
	\mathbf{B}_\mathrm{int}&=(B_\mathrm{int}\sin\theta_\mathrm{B},B_\mathrm{int}\cos\theta_\mathrm{B}),
\end{align}
where $m_\mathrm{Cu}$ = $|\mathbf{m}_\mathrm{Cu}|$. Utilizing eq. (\ref{eq:Bint}), we obtain the following relations:
\begin{align}
	&B_\mathrm{int}=m_\mathrm{Cu}(A_\parallel^2\cos^2\theta_\mathrm{m}+A_\perp^2\sin^2\theta_\mathrm{m})^{1/2},\\
	&\tan\theta_\mathrm{B}=\frac{A_\perp}{A_\parallel}\tan\theta_\mathrm{m}.
\end{align}
Here $A_\parallel$ = $A_{ZZ}$ and $A_\perp$ = $A_{X^\prime X^\prime}$ are the principal values of $\mathsf{A}$. 
They are estimated as $A_\parallel$ = $-24.1$ T/$\mathrm{\mu_B}$ and $A_\perp$ = 2.71 T/$\mathrm{\mu_B}$ for $d(x^2-y^2)$, $A_\parallel$ = 10.2 T/$\mathrm{\mu_B}$ and $A_\perp$ = $-12.2$ T/$\mathrm{\mu_B}$ for $d(3z^2-r^2)$.\cite{yoshida07}.  
Using these values, we calculated $B_\mathrm{int}$ and $\theta_\mathrm{B}$ as a function of $\theta_\mathrm{m}$ for $d(x^2-y^2)$ and $d(3z^2-r^2)$.

\begin{figure}[t]
\centerline{\includegraphics[width=86mm]{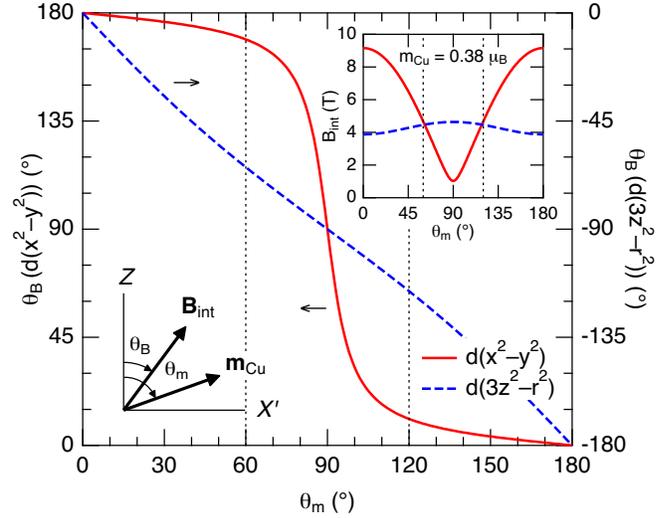}}
\caption{
(Color online) Relation between the angles $\theta_\mathrm{B}$ and $\theta_\mathrm{m}$ 
for $d(x^2-y^2)$ (solid line) and $d(3z^2-r^2)$ (dashed line) orbitals. The inset shows $B_\mathrm{int}$ as a function of $\theta_\mathrm{m}$ for the two orbitals with $m_\mathrm{Cu}$ = 0.38 $\mu_\mathrm{B}$. Dotted lines in the main panel and the inset correspond to $\theta_\mathrm{m}$ of 60$^\circ$ and 120$^\circ$. 
}
\label{fig:theta}
\end{figure}

\begin{figure}[t]
\centerline{\includegraphics[width=80mm]{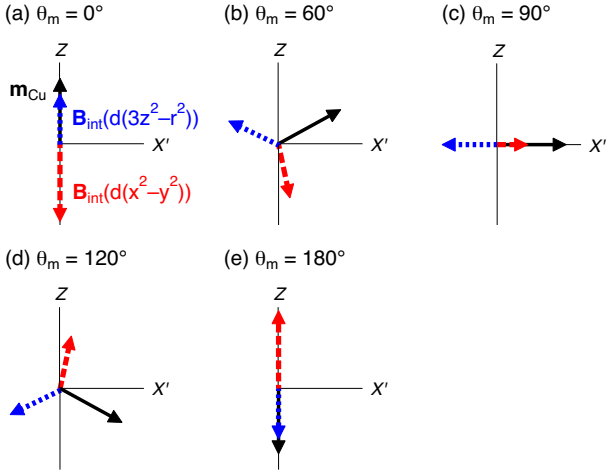}}
\caption{
(Color online) Change of the direction of the internal field at the Cu site accompanied by a rotation of the Cu magnetic moment. Solid arrow represents the Cu magnetic moment. Dashed and dotted arrows are the internal fields for $d(x^2-y^2)$ and $d(3z^2-r^2)$ orbitals, respectively. The angle $\theta_\mathrm{m}$ between the magnetic moment and the $Z$ axis is (a) 0$^\circ$, (b) 60$^\circ$, (c) 90$^\circ$, (d) 120$^\circ$, and (e) 180$^\circ$.
}
\label{fig:rotation}
\end{figure}

Figure \ref{fig:theta} shows the angle $\theta_\mathrm{B}$ plotted against $\theta_\mathrm{m}$. 
Obviously, we always have a negative slope of $\theta_\mathrm{B}$ with respect to $\theta_\mathrm{m}$, i.e., $d\theta_\mathrm{B}/d\theta_\mathrm{m}$ $<$ 0 for both the orbitals. 
In other words, $\mathbf{B}_\mathrm{int}$ rotates counterclockwise when $\mathbf{m}_\mathrm{Cu}$ rotates clockwise as illustrated in Fig. \ref{fig:rotation}. This comes from the fact that $A_\parallel$ and $A_\perp$ are different in sign. 
The dependence of $\theta_\mathrm{B}$ on $\theta_\mathrm{m}$ is, however, qualitatively very different between the two orbitals. 

We examine the result for $d(x^2-y^2)$ first. 
When $\mathbf{m}_\mathrm{Cu}$ is directed along the $Z$ axis, i.e., $\theta_\mathrm{m}$ = 0$^\circ$, $\mathbf{B}_\mathrm{int}$ also lies along the $Z$ direction, being antiparallel to $\mathbf{m}_\mathrm{Cu}$ with $\theta_\mathrm{B}$ = 180$^\circ$ (Fig. \ref{fig:rotation}(a)). This reflects the fact that $A_\parallel<0$ for $d(x^2-y^2)$. 
The angle $\theta_\mathrm{B}$ decreases gradually with increasing $\theta_\mathrm{m}$, 
yet the resultant canting of $\mathbf{B}_\mathrm{int}$ from the $Z$ direction remains small, being only within 11$^\circ$ ($\theta_\mathrm{B}\geq 169^\circ$) for $\theta_\mathrm{m}$ up to 60$^\circ$. 
Significant change of $\theta_\mathrm{B}$ occurs in a range $80^\circ$ $\lesssim$ $\theta_\mathrm{m}$ $\lesssim$ $100^\circ$  
where more than 110$^\circ$ change of $\theta_\mathrm{B}$ takes place due to only 20$^\circ$ change of $\theta_\mathrm{m}$.  
$\mathbf{m}_\mathrm{Cu}$ and $\mathbf{B}_\mathrm{int}$ become collinear again along the $X^\prime$ axis with $\theta_\mathrm{m}$ = $\theta_\mathrm{B}$ = 90$^\circ$ (Fig. \ref{fig:rotation}(c)). 

The direction of $\mathbf{B}_\mathrm{int}$ for $d(3z^2-r^2)$ changes more smoothly with $\theta_\mathrm{m}$. Since $A_\parallel>0$, $\mathbf{m}_\mathrm{Cu}$ and $\mathbf{B}_\mathrm{int}$ are parallel to each other when $\theta_\mathrm{m}$ = 0$^\circ$. $\theta_\mathrm{B}$ varies almost linearly with increasing $\theta_\mathrm{m}$, satisfying the approximate relation $\theta_\mathrm{B}\approx -\theta_\mathrm{m}$. 
Along the $X^\prime$ axis, $\mathbf{B}_\mathrm{int}$ is oriented antiparallel to $\mathbf{m}_\mathrm{Cu}$. 

When $\theta_\mathrm{m}$ = 60$^\circ$ or 120$^\circ$ which is the case in Cu$_2$Fe$_2$Ge$_4$O$_{13}$, we have $\theta_\mathrm{B}$ = 169$^\circ$ or 11$^\circ$ for $d(x^2-y^2)$. 
It should be noted here that there is no distinction between the two sets of the angles $(\theta_\mathrm{m},\theta_\mathrm{B})=(60^\circ,169^\circ)$ and $(120^\circ,11^\circ)$: 
they are equivalent because of twofold symmetry about the $X^\prime$ axis. 
See Figs. \ref{fig:rotation}(b) and \ref{fig:rotation}(d). 
In either case the deviation of $\mathbf{B}_\mathrm{int}$ from the $Z$ direction is only 11$^\circ$ despite the large canting angle (60$^\circ$) of $\mathbf{m}_\mathrm{Cu}$. 
On the other hand, $\mathbf{B}_\mathrm{int}$ for $d(3z^2-r^2)$ is directed away from the $Z$ direction to have a large canting angle of 64$^\circ$ ($\theta_\mathrm{B}$ = $-64^\circ$ or $-116^\circ$).  

The dependence of $B_\mathrm{int}$ on $\theta_\mathrm{m}$ is also very contrasting between the two orbitals as shown in the inset of Fig. \ref{fig:theta}. $B_\mathrm{int}$ for $d(x^2-y^2)$ depends strongly on $\theta_\mathrm{m}$: for $m_\mathrm{Cu}$ = 0.38 $\mu_\mathrm{B}$, it varies from 9 to 1 T as $\theta_\mathrm{m}$ varies from 0$^\circ$ to 90$^\circ$. In contrast, $B_\mathrm{int}$ for $d(3z^2-r^2)$ changes only slightly with $\theta_\mathrm{m}$, ranging from 3.9 to 4.6 T. When $\theta_\mathrm{m}$ = 60$^\circ$ and 120$^\circ$, two orbitals give accidentally the same value $B_\mathrm{int}$ = 4.6 T. This is not very different from the observed value of 3.39 T. 
Notice also that the large canting angle of $\mathbf{m}_\mathrm{Cu}$ as well as the reduced value of $m_\mathrm{Cu}$ is responsible for the modestly large value of $B_\mathrm{int}$ for $d(x^2-y^2)$. 

It is apparent from the above discussion that anisotropy of the hyperfine interaction of the $d(x^2-y^2)$ orbital accounts qualitatively for the small canting angle of $\mathbf{B}_\mathrm{int}$ which arises from $\mathbf{m}_\mathrm{Cu}$ 
being tilted away from the symmetry ($Z$) axis. 
This suggests a dominant $d(x^2-y^2)$ character of the Cu$^{2+}$ ions in Cu$_2$Fe$_2$Ge$_4$O$_{13}$. 
The $d(x^2-y^2)$ character would be understood by remembering the fourfold planar coordination of Cu. The fourfold, square planar coordination of Cu by O is derived from a CuO$_6$ octahedron by removing the apical oxygens. The point symmetry at the Cu site is tetragonal but the $d(x^2-y^2)$ orbital is energetically favored because there is no apical oxygen. Note also that this orbital extends to the directions of Cu-O bonding, allowing the Cu$^{2+}$ ion to exchange interact via the O$^{2-}$ ions with the neighboring Cu$^{2+}$ ion which shares a common edge of the CuO$_4$ squares. The $d(x^2-y^2)$ nature of the ground state orbital seems thus compatible with the strong intradimer interactions observed in Cu$_2$Fe$_2$Ge$_4$O$_{13}$ and Cu$_2$Sc$_2$Ge$_4$O$_{13}$.

There are some quantitative discrepancies between the experimental results and the model calculations, especially in the magnitude $B_\mathrm{int}$ of the internal field.  This may be attributed to mixing of the $d(3z^2-r^2)$ orbital to the ground state allowed by distortion from the ideal square planar geometry which is manifested by a moderate asymmetry $\eta=0.283$ of the EFG tensor. A preliminary model calculation shows that 10 to 20 \% mixing of the $d(3z^2-r^2)$ orbital improves dramatically the quantitative discrepancies, yielding a canting angle of $\mathbf{B}_\mathrm{int}$ within 6$^\circ$ from the $Z$ axis and $B_\mathrm{int}$ of 3.3 to 3.9 T.

\subsection{Nuclear spin-lattice relaxation at high temperatures} 

Nuclear spin-lattice relaxation rate is one of the most important quantities measured by NMR and NQR, because it reflects low-energy magnetic excitations quite sensitively. 
Here we discuss the nuclear spin-lattice relaxation at the Cu site at high temperatures in both Cu$_2$Sc$_2$Ge$_4$O$_{13}$ and Cu$_2$Fe$_2$Ge$_4$O$_{13}$. 

First we discuss the relaxation rate in Cu$_2$Sc$_2$Ge$_4$O$_{13}$. In the system with localized magnetic moments, the nuclear spin-lattice relaxation rate of $I$ = $3/2$ nuclei measured by NQR is given as\cite{moriya56,chepin91}
\begin{align}
	\frac{1}{T_\mathrm{1\infty,NQR}}=&\sqrt{\frac{\pi}{2}}\frac{\gamma^2 S(S+1)}{3(3+\eta^2)\omega_\mathrm{ex}}\nonumber\\
	&\times\left[(3+\eta)^2 A_{XX}^2+(3-\eta)^2 A_{YY}^2+4\eta^2 A_{ZZ}^2\right]\label{eq:T1}
\end{align}
in the high-temperature limit. 
Here $S$ is the electronic spin, $A_{\mu\mu}$'s $(\mu=X,Y,Z)$ are the principal values of the hyperfine tensor. $\omega_\mathrm{ex}$ is the exchange frequency which is given for an isolated spin dimer as 
\begin{align}
	\omega_\mathrm{ex}^2=\frac{2}{3}S(S+1)\left(\frac{J_0}{\hbar}\right)^2\label{eq:wex0},
\end{align}
where $J_0$ is the intradimer exchange interaction.
If we assume axial symmetry of the hyperfine tensor $A_{XX}=A_{YY}=A_\perp$ and $A_{ZZ}=A_\parallel$ for simplicity, eq. (\ref{eq:T1}) reduces to
\begin{align}
	\frac{1}{T_\mathrm{1\infty,NQR}}&=\sqrt{\frac{\pi}{2}}\frac{2\gamma^2 A_\perp^2 S(S+1)}{3\omega_\mathrm{ex}}f(\eta,r),\nonumber\\
	f(\eta,r)&=3+\frac{2\eta^2 (r^2-1)}{3+\eta^2}.\label{eq:T1axial}
\end{align}
Here $r$ = $A_\parallel/A_\perp$, and the function $f(\eta,r)$ describes effects of asymmetry $\eta$ of the EFG and the anisotropy $r$ of the hyperfine coupling. 
Putting $S$ = $1/2$, $J_0$ = 24 meV into eq. (\ref{eq:wex0}), we obtain $\omega_\mathrm{ex} = 2.6\times 10^{13}$ s$^{-1}$. If we take the $d(x^2-y^2)$ orbital state for a Cu$^{2+}$ ion, 
we have $r$ = $-8.89$.  
We then obtain $1/T_\mathrm{1\infty,NQR}$ = $2.5\times 10^4$ s$^{-1}$ for $^{63}$Cu from eq. (\ref{eq:T1axial}) using $\eta$ = 0.283. Note that for $\eta$ = 0.283 and $r$ = $-8.89$, $f(\eta,r)$ = 7.06 which is much larger than the value of the function for cases with $\eta$ = 0 and/or $r$ = 1. This value of $1/T_\mathrm{1\infty,NQR}$ seems to be in the right orders of magnitude because $1/T_\mathrm{1,NQR}\mathrm{(Sc)}$ is expected to reach around a value of $10^5$ s$^{-1}$ in the high-temperature limit judging from its temperature dependence shown in Fig. \ref{fig:T1_actv}.

\begin{figure}[t]
\centerline{\includegraphics[width=42mm]{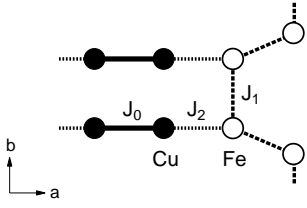}}
\caption{Local arrangement of magnetic atoms around Cu and definition of the exchange interactions $J_0$, $J_1$ and $J_2$.}
\label{fig:exchange}
\end{figure}

In Cu$_2$Fe$_2$Ge$_4$O$_{13}$, we expect a contribution of the neighboring Fe$^{3+}$ spin to $1/T_\mathrm{1\infty,NQR}$ at the Cu site via the transferred hyperfine interaction in addition to the on-site one represented by eq. (\ref{eq:T1}) or (\ref{eq:T1axial}). $1/T_\mathrm{1\infty,NQR}$ at the Cu site in Cu$_2$Fe$_2$Ge$_4$O$_{13}$ is hence expressed as
\begin{align}
	\frac{1}{T_\mathrm{1\infty,NQR}}&=\left(\frac{1}{T_\mathrm{1\infty,NQR}}\right)_\mathrm{Cu}+\left(\frac{1}{T_\mathrm{1\infty,NQR}}\right)_\mathrm{Fe},
\end{align}
where the first and second terms are the contributions of Cu$^{2+}$ and Fe$^{3+}$ spins, respectively. 
Assuming the isotropic transferred hyperfine interaction with a coupling constant $B$ from the nearest neighbor Fe$^{3+}$ ion, we have
\begin{align}
	\left(\frac{1}{T_\mathrm{1\infty,NQR}}\right)_\mathrm{Cu}&=\sqrt{\frac{\pi}{2}}\frac{2\gamma^2 A_\perp^2 S_\mathrm{Cu}(S_\mathrm{Cu}+1)}{3\omega_\mathrm{ex,Cu}}f(\eta,r),\label{eq:T1Cu}\\
	\left(\frac{1}{T_\mathrm{1\infty,NQR}}\right)_\mathrm{Fe}&=3\sqrt{\frac{\pi}{2}}\frac{2\gamma^2 B^2 S_\mathrm{Fe}(S_\mathrm{Fe}+1)}{3\omega_\mathrm{ex,Fe}}\label{eq:T1Fe}.
\end{align}
Here $S_\mathrm{Cu}$ = $1/2$ ($S_\mathrm{Fe}$ = $5/2$) is the electronic spin of a Cu$^{2+}$ (Fe$^{3+}$) ion.
Note that a factor 3 in the right hand side of eq. (\ref{eq:T1Fe}) comes from the fact that we are dealing with NQR. The exchange frequency $\omega_{\mathrm{ex},i}$ of the electronic spin at the $i$-th site is generally given as
\begin{align}
	\omega_{\mathrm{ex},i}^2=\frac{2}{3}\sum_jS_j(S_j+1)\left(\frac{J_{ij}}{\hbar}\right)^2,\label{eq:wex1}
\end{align}
where $S_j$ is the electronic spin at the $j$-th site and $J_{ij}$ is the exchange interaction between the $i$-th and $j$-th spins. Applying eq. (\ref{eq:wex1}) to the Cu$^{2+}$ and Fe$^{3+}$ ions in Cu$_2$Fe$_2$Ge$_4$O$_{13}$, we have
\begin{align}
	\omega_\mathrm{ex,Cu}^2&=\frac{2}{3}\left[\frac{3}{4}\left(\frac{J_0}{\hbar}\right)^2+\frac{35}{4}\left(\frac{J_2}{\hbar}\right)^2\right],\label{eq:wex2}\\
	\omega_\mathrm{ex,Fe}^2&=\frac{2}{3}\left[2\cdot\frac{35}{4}\left(\frac{J_1}{\hbar}\right)^2+\frac{3}{4}\left(\frac{J_2}{\hbar}\right)^2\right].\end{align}
The exchange interactions $J_0$, $J_1$ and $J_2$ are defined as shown in Fig. \ref{fig:exchange}. We neglected the interchain Fe-Fe interaction ($J_\mathrm{Fe}^\prime$) in the $c$ direction because it is an order of magnitude smaller than $J_1$ and $J_2$. Using the values $J_0$ = $J_\mathrm{Cu}$ = 24 meV, $J_1$ = $J_\mathrm{Fe}$ = 1.60 meV, $J_2$ = $J_\mathrm{Cu-Fe}$ = 2.54 meV determined by the inelastic neutron scattering experiments,\cite{masuda05} we obtain $\omega_\mathrm{ex,Cu}$ = $2.7\times 10^{13}$ s$^{-1}$ and $\omega_\mathrm{ex,Fe}$ = $8.7\times 10^{12}$ s$^{-1}$.

It is reasonable to consider the temperature-independent contribution of Fe$^{3+}$ spins to $1/T_\mathrm{1,NQR}(\mathrm{Fe})$, i.e., the term $a$ of eq. (\ref{eq:T1scale}), as $(1/T_\mathrm{1\infty,NQR})_\mathrm{Fe}$. By equating $(1/T_\mathrm{1\infty,NQR})_\mathrm{Fe}$ to $a$ = $2.67\times 10^3$ s$^{-1}$, the only unknown quantity in eq. (\ref{eq:T1Fe}), the absolute value of the transferred hyperfine coupling $B$, is estimated as $|B|$ = 0.23 T/$\mu_\mathrm{B}$. This is an order of magnitude larger than a classical dipole field from the surrounding Fe$^{3+}$ ions and is in a reasonable range for the transferred hyperfine interaction.\cite{fn:transfer} 

The on-site contribution $(1/T_\mathrm{1\infty,NQR})_\mathrm{Cu}$ is estimated to be $2.4\times 10^4$ s$^{-1}$ using $\omega_\mathrm{ex,Cu}$ calculated above and assuming the $d(x^2-y^2)$ orbital state. 
This is in good agreement with a value $(1/T_\mathrm{1\infty,NQR}\mathrm{(Fe)})_\mathrm{Cu}$ $\sim$ $2$ $\times$ $10^4$ s$^{-1}$ evaluated by extrapolating the temperature dependence of $(1/T_\mathrm{1,NQR}\mathrm{(Fe)})_\mathrm{Cu}$ to $T$ $\rightarrow$ $\infty$ utilizing Fig. \ref{fig:T1_actv}. 

Lastly we discuss the numerical factor $b$ in eq. (\ref{eq:T1scale}) relating $1/T_{1,\mathrm{NQR}}(\mathrm{Sc})$ to $(1/T_{1,\mathrm{NQR}}(\mathrm{Fe}))_\mathrm{Cu}$. 
Since the temperature dependences of $1/T_{1,\mathrm{NQR}}(\mathrm{Sc})$ and $(1/T_{1,\mathrm{NQR}}(\mathrm{Fe}))_\mathrm{Cu}$ are identical apart from the factor $b$, we may argue their values in the high-temperature limit using eqs. (\ref{eq:T1axial}) and (\ref{eq:T1Cu}). The difference of the two quantities then comes primarily from the difference of the hyperfine coupling constant $A_\perp$ between the two compounds, because the exchange frequencies of the Cu$^{2+}$ spin given by eqs. (\ref{eq:wex0}) and (\ref{eq:wex2}) are nearly the same. Utilizing the fact that $1/T_\mathrm{1\infty,NQR}$ $\propto$ $A_\perp^2$ and the obtained value of $b$, we evaluated a ratio of $A_\perp$ between the Fe and Sc compounds as $A_\perp\mathrm{(Fe)}/A_\perp\mathrm{(Sc)}$ $\sim b^{1/2}$ $\approx$ 0.46. 

It is known that a small and positive value of $A_\perp$ for the $d(x^2-y^2)$ orbital results from subtle balance of core polarization, dipole and orbital fields which are different in sign and are an order of magnitude larger than the total $A_\perp$\cite{AB}. 
On the one hand, we have in Cu$_2$Fe$_2$Ge$_4$O$_{13}$ the exchange interaction between Cu and Fe atoms which is absent in Cu$_2$Sc$_2$Ge$_4$O$_{13}$. 
It is thus probable that a microscopic process relevant to the Cu-Fe interaction modifies a wave function of the Cu atom in Cu$_2$Fe$_2$Ge$_4$O$_{13}$ from that in Cu$_2$Sc$_2$Ge$_4$O$_{13}$, 
altering the balance of various contributions to $A_\perp$. 
The factor of $\sim$2 difference of $A_\perp$ between the two compounds might be attributed to such a change of the electronic state of a Cu$^{2+}$ ion.

\section{Summary}
\label{sec:summary}

We have presented the results of Cu NQR and NMR in a bicomponent magnet Cu$_2$Fe$_2$Ge$_4$O$_{13}$ including quantum spin dimers and classical spin chains. 
The results of Cu NQR in a reference compound Cu$_2$Sc$_2$Ge$_4$O$_{13}$ containing only spin dimers have also been reported. 

In the paramagnetic state of Cu$_2$Fe$_2$Ge$_4$O$_{13}$, the $^{63}$Cu nuclear spin-lattice relaxation rate $1/T_\mathrm{1,NQR}$ has two independent contributions. 
One is from Fe$^{3+}$ spins which is independent of temperature, and the other is due to Cu$^{2+}$ spins exhibiting a thermally-activated behavior except near $T_\mathrm{N}$. The temperature dependence of the latter is described by a spin gap identical to that in Cu$_2$Sc$_2$Ge$_4$O$_{13}$.  
Such a characteristic of $1/T_\mathrm{1,NQR}$ suggests that the Fe chains contribute almost independently to the magnetic excitation spectrum without affecting the gapped excitations of the Cu dimers. This is consistent with the results of the neutron scattering experiments and should be ascribed to different energy scales between the Cu dimers and the Fe chains. 

There appears an internal field at the Cu site in the antiferromagnetic state accompanied by ordering of Cu magnetic moments. 
The internal field is nearly parallel to the $Z$ principal axis of the EFG tensor, but is directed away from the ordered moment of the Cu atom. For this property of the internal field, we introduced a simplified model of the internal field based on the fourfold planar coordination of Cu. 
The model revealed that the relation between the directions of the internal field, ordered moment and the $Z$ principal axis of the EFG tensor is well described by an anisotropic hyperfine interaction of the $d(x^2-y^2)$ orbital. 
The $d(x^2-y^2)$ orbital is compatible with a modestly large value of the internal field 
if we take account of a large canting angle of the ordered moment from the $Z$ axis as well as a reduction of the moment from a fully polarized value.    
The $d(x^2-y^2)$ orbital also predicts a value of $1/T_\mathrm{1,NQR}$ in the high-temperature limit which agrees well with that evaluated from the experimental data. 
We therefore conclude that the ground state orbital of a Cu$^{2+}$ ion in Cu$_2$Fe$_2$Ge$_4$O$_{13}$ has predominantly a $d(x^2-y^2)$ character.

\section*{Acknowledgment}
\begin{acknowledgment}

This work was partly supported by Grant-in-Aid for Scientific Research (C) 22540352 from the Japan Society for the Promotion of Science (JSPS).


\end{acknowledgment}

%
%
%


\begin{thebibliography}{99}

\bibitem{zheludev98} A. Zheludev, E. Ressouche, S. Maslov, T. Yokoo, S. Raymond and J. Akimitsu: Phys. Rev. Lett. \textbf{80} (1998) 3630.

\bibitem{masuda04} T. Masuda, A. Zheludev, B. Grenier, S. Imai, K. Uchinokura, E. Ressouche and S. Park: Phys. Rev. Lett. \textbf{93} (2004) 077202.

\bibitem{hase05} M. Hase, M. Konho, H. Kitazawa, O. Suzuki, K. Ozawa, G. Kido, M. Imai and X. Hu: Phys. Rev. B \textbf{72} (2005) 
172412.

\bibitem{hamasaki08} T. Hamasaki, T. Ide, H. Kuroe, T. Sekine, M. Hase, I. Tsukada and T. Sakakibara: Phys. Rev. B \textbf{77} (2008) 134419.

\bibitem{masuda03} T. Masuda, B. C. Chakoumakos, C. L. Nygren, S. Imai and K. Uchinokura: J. Solid State Chem. \textbf{176} (2003) 175.

\bibitem{hase93} M. Hase, I. Terasaki and K. Uchinokura: Phys. Rev. Lett. \textbf{70} (1993) 3651.

\bibitem{masuda05} T. Masuda and A. Zheludev, B. Sales, S. Imai, K. Uchinokura and S. Park: Phys. Rev. B \textbf{72} (2005) 094434.

\bibitem{masuda07} T. Masuda, K. Kakurai and M. Matsuda, K. Kaneko and N. Metoki: Phys. Rev. B \textbf{75} (2007) 220401. 

\bibitem{matsumoto10a} M. Matsumoto, H. Kuroe, T. Sekine and T. Masuda: J. Phys. Soc. Jpn. \textbf{79} (2010) 084703.

\bibitem{matsumoto10b} M. Matsumoto, H. Kuroe, T. Sekine and T. Masuda: J. Phys: Conference Series \textbf{200} (2010) 022034.

\bibitem{redhammer04} G. J. Redhammer and G. Roth: J. Solid State Chem. \textbf{177} (2004) 2714.

\bibitem{masuda06} T. Masuda and G. J. Redhammer: Phys. Rev. B \textbf{74} (2006) 054418.

\bibitem{lue07} C. S. Lue, C. N. Kuo, T. H. Su and G. J. Redhammer: Phys. Rev. B \textbf{75} (2007) 014426.

\bibitem{jkiku10} J. Kikuchi, S. Nagura, H. Nakanishi and T. Masuda: J. Phys: Conference Series \textbf{200} (2010) 022024.

\bibitem{clark95} W. G. Clark, M. E. Hanson, F. Lefloch and P. S\'{e}gransan: Rev. Sci. Instrum. \textbf{66} (1995) 2453.

\bibitem{shimizu93} T. Shimizu: J. Phys. Soc. Jpn. \textbf{62} (1993) 772.

\bibitem{mchenry72} M. R. McHenry, B. G. Silbernagel and J. H. Wernick: Phys. Rev. B \textbf{5} (1972) 2958.

\bibitem{fn:perturbation} Magnitude of the Zeeman interaction $|\mathcal{H}_Z|$ may roughly be evaluated from the frequency of the central transition (37.8 MHz and 40.8 MHz for $^{63}$Cu and $^{65}$Cu, respectively), whereas the NQR frequency  (34.90 MHz and 32.29 MHz for $^{63}$Cu and $^{65}$Cu at 50 K) would give an estimate of the quadrupole interaction strength $|\mathcal{H}_Q|$.

\bibitem{abragam} A. Abragam: \textit{Principles of Nuclear Magnetism} (Oxford University Press, Oxford, 1961).

\bibitem{fn:Delta} For a given value of $\eta$, the solutions of $\Delta(\theta)=\Delta_\mathrm{obs}$ are different between $^{63}$Cu and $^{65}$Cu. This is due to the fact that the second-order perturbation theory does not give accurately the quadruple splitting.

\bibitem{fn:twofold} Since the Hamiltonian eq. (\ref{eq:Hamil}) has twofold symmetry about the EFG principal axes, it is sufficient to take the ranges $0\leq\theta\leq 90^\circ$ and $0\leq\phi\leq 90^\circ$ into account.

\bibitem{fn:prob} All the matrix elements of $I_X$, $I_Y$ and $I_Z$ corresponding to the lines $d$ and $d^\prime$ increase with increasing $\phi$. Therefore, a tendency of the lines $d$ and $d^\prime$ becoming more intense as $\phi$ is varied from $0^\circ$ to $90^\circ$ does not depend on the fact that we take a powder average of $I_{\alpha\beta}$.

\bibitem{yoshida07} M. Yoshida, N. Ogata, M. Takigawa, J. Yamaura, M. Ichihara, T. Kitano, H. Kageyama, Y. Ajiro and K. Yoshimura: J. Phys. Soc. Jpn. \textbf{76} (2007) 104703.

\bibitem{moriya56} T. Moriya: Prog. Theor. Phys. \textbf{16} (1956) 641.

\bibitem{chepin91} J. Chepin and J. H. Ross, Jr.: J. Phys.: Condens. Matter \textbf{3} (1991) 8103.

\bibitem{fn:transfer} The transferred hyperfine interaction produces the internal field at the Cu site of about 0.8 T due to the ordered moment of an Fe atom. This is much smaller than the internal field arising from the on-site interaction, so that the direction of the internal field does not change a lot. 

\bibitem{AB} A. Abragam and B. Bleaney: \textit{Electron Paramagnetic Resonance of Transition Ions} (Oxford University Press, Oxford, 1970).

\end{thebibliography}
\end{document}